\documentclass[fleqn,12pt,twoside]{article}
\usepackage{espcrc1}

\usepackage{graphicx}

\title{Summary of Results on Hard Probes}

\author{T. Peitzmann \address{University of M\"unster, 48149 M\"unster, Germany} 
\thanks{now at: GSI, 64291 Darmstadt, Germany}}
       
\begin{document}

\maketitle

\begin{abstract}
Results on observables related to hard
scattering processes as presented at this conference are reviewed. 
This includes in particular measurements
related to jet quenching, a phenomenon which has been predicted as a 
signature for the hot and dense early phase of strongly interacting
matter.
\end{abstract}

\section{INTRODUCTION}
With the new experiments at the RHIC accelerator at BNL, experimental
heavy ion physics has truly reached the domain of hard probes. At
RHIC energies hard scatterings as observed via jet structures are
important in p+p collisions, and it was 
expected that hard scatterings should also be an important feature of
heavy ion collisions at the collider. 
It has been predicted \cite{aaa1} that hard scattered partons
should suffer energy loss from medium induced gluon radiation in a
hot and dense system like the expected quark-gluon plasma phase of
strongly interacting matter. This ``jet-quenching'', which will lead 
to a suppression of hadron production at large transverse momenta, is therefore
considered a signature of the partonic initial state. (For a review
see \cite{aaa1a}.)

Two main avenues are being
followed to get access to hard scattering processes:
\begin{itemize}
    \item  The study of high $p_{T}$ particle production via singles 
    spectra and their comparison for different centralities and
    reaction systems (including p+p collisions).

    \item  The study of azimuthal correlations reminiscent of jet
    topologies.
\end{itemize}
At the last Quark Matter conference \cite{aaa2} a suppression of high $p_{T}$ 
hadron production has already been reported \cite{aaa3}, and the
results have been published by the PHENIX and STAR experiments \cite{aaa4,aab4}.

At this conference, new results from all four RHIC experiments have
been reported. The BRAHMS \cite{aaa5} and PHOBOS \cite{aaa6} 
experiments have presented spectra of hadrons out to relatively 
high $p_{T}$, while PHENIX\cite{aaa7} and STAR \cite{aaa8} have shown
truly impressive results on hadrons out to very high $p_{T}$. I will
try to summarize the highlights in the following.

\section{SCALING OF PARTICLE PRODUCTION}
The role of hard scattering even in understanding the global
observables has been discussed in a number of papers \cite{bbb1},
where it has been suggested that, in studying the centrality
dependence of the charged particle pseudorapidity density
$dN_{ch}/d\eta$, one might
extract information on the relative contribution from hard
scatterings. In particular, a parameterization as a function of the
number of participants and the number of collisions has been proposed:
\begin{equation}
    \frac{dN_{ch}}{d\eta} = A \cdot N_{part} + B \cdot N_{coll},
    \label{eq:b1}
\end{equation}
where the necessity of a term proportional to the number of collisions
to explain the experimental data
was taken as a hint for hard scattering production.
In this context, the investigation of the scaling properties of
particle production as presented by the PHOBOS experiment \cite{bbb2} is
interesting. PHOBOS reported that the total charged multiplicity scales
in good approximation with the number of participants and no
additional term proportional to the number of collisions is necessary.
The produced particles are however distributed differently in
rapidity for varying centrality. This possibly implies that the
observed stronger variation of $dN_{ch}/d\eta$ with centrality may be
due to a measurement in a limited phase space region combined with a 
redistribution of particles in phase space. In light of these
observations, the interpretation of the parameterization
(\ref{eq:b1}) 
as related to hard scattering appears to be premature.

\section{PRODUCTION OF HIGH $\mathbf{P_T}$ INCLUSIVE HADRONS}

Momentum spectra of hadrons have been reported by all four RHIC
experiments. Qualitatively, the suppression reported from the
measurements at $\sqrt{s_{NN}} = 130 \, \mathrm{GeV}$ 
has also been found at $\sqrt{s_{NN}} = 200 \, \mathrm{GeV}$. Charged
hadrons have been measured out to $p_{T} = 9 \, \mathrm{GeV}/c$ by
PHENIX \cite{ccc1} and to $p_{T} = 11 \, \mathrm{GeV}/c$ by
STAR \cite{ccc2}. 
Identified hadrons at very high $p_{T}$, which do not suffer from
uncertainties due to the unknown flavour composition, have only
been presented by the PHENIX experiment \cite{ccc3}, 
which has measured neutral pion 
spectra out to $p_{T} = 10 \, \mathrm{GeV}/c$. 

\begin{figure}[tb]
    \centering
    \includegraphics{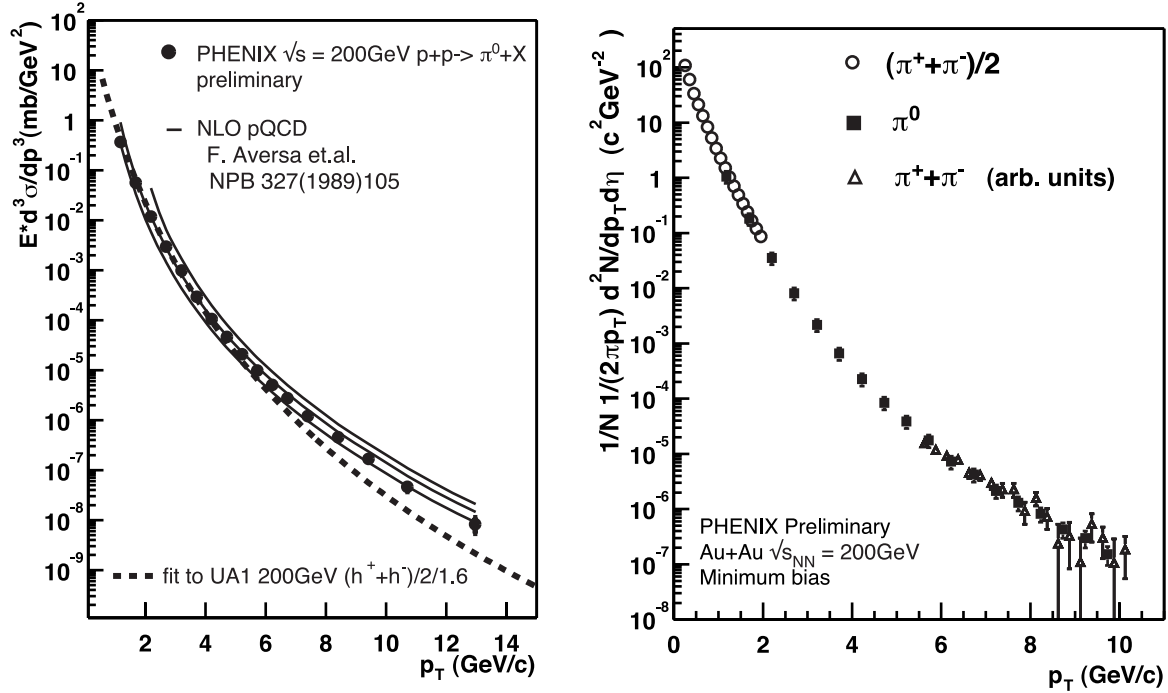}
    \caption{Left: Neutral pion spectra for p+p collisions at 200 GeV 
    measured by the PHENIX experiment \protect\cite{ccc7}. 
    Right: Pion spectra for
    minimum bias Au+Au collisions at $\sqrt{s_{NN}} = 200 \,
    \mathrm{GeV}$ 
    measured by the
    PHENIX experiment. In addition to the neutral pions
    charged pions identified by TOF at low $p_{T}$ and by RICH at
    high $p_{T}$ are included \protect\cite{aaa7}.}
    \label{fig1}
\end{figure}

To quantitatively judge the measured hadron yield in central Au+Au
collisions one tries to compare it to the expectations from p+p
collisions, which must be appropriately scaled to the larger systems.
As hard scatterings have a very small cross section and are expected 
to be incoherent, it is reasonable to assume that the multiplicity should
scale with the number of binary nucleon-nucleon collisions. In fact, 
the scaling of high $p_{T}$ hadron production in p+A collisions at
somewhat lower energies has been shown to be stronger than with the
number of collisions (``Cronin-effect'') \cite{ccc4}. A qualitatively
similar enhancement compared to p+p collisions was found in 
central Pb+Pb collisions at 17.3~$A$GeV for neutral pion production 
by the WA98 experiment \cite{ccc5}. In the light of this known
enhancement at lower energies, the suppression observed at RHIC is
even more striking.

It has become customary to illustrate the suppression by using the
so-called nuclear modification factor:
\begin{equation}
    R_{AA}(p_{T}) \equiv 
    \frac{d^{2}N^{A+A} / dp_T d\eta}  
    {\langle N_{coll} \rangle
    d^{2}N^{N+N} / dp_T d\eta}.
    \label{eq:raa}
\end{equation}

In addition to the measurement of the hadron spectrum in heavy ion
collisions this requires the knowledge of
\begin{enumerate}
    \item[(1)]  the number of binary collisions and 

    \item[(2)]  the spectrum in nucleon-nucleon collisions (or p+p 
    collisions) at the same energy.
\end{enumerate}
(1) is usually obtained by calculating the nuclear overlap in a
Glauber-like approach. The systematic errors involved are among
others related to uncertainties in the nucleon-nucleon cross section, 
the nuclear density profile and the trigger bias in the heavy ion
measurement. While for peripheral collisions these do result in
considerable uncertainties, the systematic error of the number of
binary collisions $N_{coll}$ in central collisions is estimated to be
at most $10 \% $. 
As a nucleon-nucleon spectrum (2) up to now data from earlier experiments at
different energies were extrapolated to the RHIC conditions. At
$\sqrt{s_{NN}} = 200 \, \mathrm{GeV}$ 
charged hadron spectra from UA1 \cite{ccc6} in
$\mathrm{\bar{p}+p}$-collisions are available up to 
$p_{T} = 6 \, \mathrm{GeV}/c$. The small momentum range and the
possibly different systematic errors due to the measurement in a
different experiment do however limit the usefulness of these data.

In this situation it is extremely important that PHENIX has measured 
the neutral pion spectra in p+p collisions at the same energy 
in the same apparatus \cite{ccc7}. This reduces the systematic
uncertainties of the comparison (i.e. in $R_{AA}$). Together with the
fact that the neutral pions are the only identified hadrons at high
$p_{T}$ this makes their measurement superior for the study of hadron
suppression compared to the charged hadrons.

The left hand side of Fig.~\ref{fig1} shows neutral pion spectra from p+p 
collisions. They agree relatively well with estimates from pQCD
calculations \cite{ccc8}. For comparison a parameterization based on the
UA1 data is also shown, which agrees with the neutral pions at low
$p_{T}$ but starts to deviate for $p_{T} > 6 \, \mathrm{GeV}/c$. The 
right hand side of Fig.~\ref{fig1} shows pion spectra in minimum bias
Au+Au collisions. In the PHENIX experiment charged pions can be identified 
by time-of-flight at low momenta. They compare nicely with the
neutral pions, which extend out to much higher momenta. In addition,
at very high $p_{T}$ there are first attempts to identify charged
pions in the RICH -- the (scaled) spectra show also nice agreement in
shape with the neutral pions.

\begin{figure}[tb]
    \centering
    \includegraphics{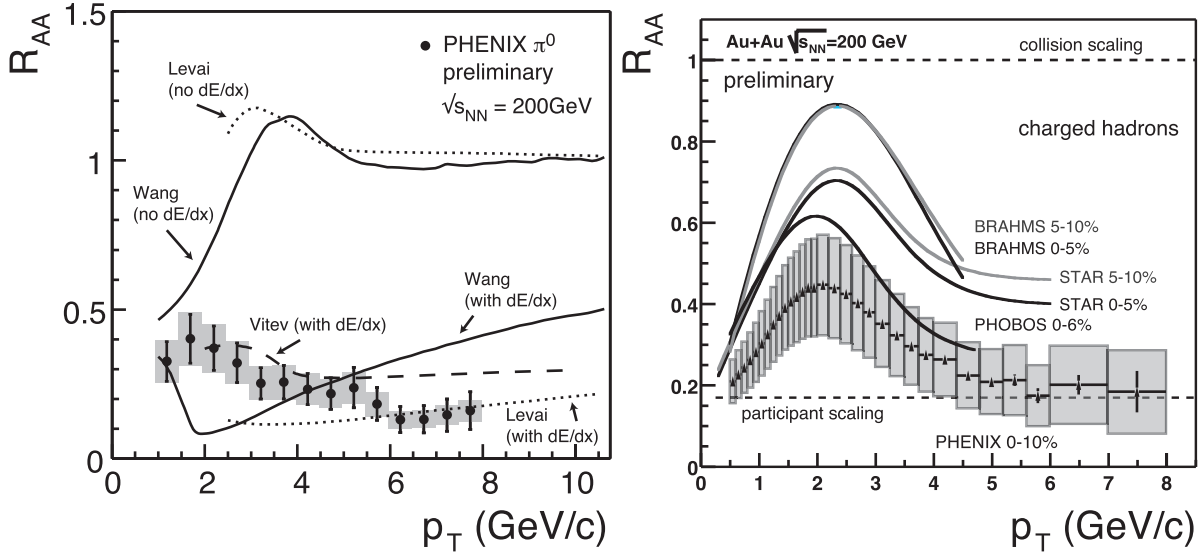}
    \caption{Left: Nuclear modification factor for neutral pions in the $10
   \% $ most central
    Au+Au collisions at $\sqrt{s_{NN}} = 200 \, \mathrm{GeV}$ 
    as presented by PHENIX
    \protect\cite{aaa7}. The curves show theoretical
    calculations without and with parton energy loss
    \protect\cite{ccc9,ccc10,ccc11}. Right: Nuclear modification
    factor for charged hadrons. The PHENIX results 
    \protect\cite{ccc1} are shown as
    symbols with shaded areas as systematic errors, the results for
    STAR \protect\cite{ccc2}, PHOBOS \protect\cite{aaa6} and 
    BRAHMS \protect\cite{aaa5} have been extracted from the
    presentations. The three latter have systematic errors of similar 
    size as the PHENIX result.}
    \label{fig2}
\end{figure}

Fig.~\ref{fig2} (left) shows the nuclear modification factor (see equation
\ref{eq:raa}) for the $10 \% $ most central collisions of Au+Au at
$\sqrt{s_{NN}} = 200 \, \mathrm{GeV}$ 
as presented by the PHENIX experiment. $R_{AA}$ is well
below one for all of the covered $p_{T}$ range, decreasing from
$\approx 0.4$ at $p_{T} = 1 - 2 \, \mathrm{GeV}/c$ to less than 0.2
for $p_{T} > 6 \, \mathrm{GeV}/c$. The values are consistent with the
results published for $\sqrt{s_{NN}} = 130 \, \mathrm{GeV}$ \cite{aaa4} 
for a much smaller $p_{T}$ range.

It is not surprising that $R_{AA} < 1$ at low $p_{T}$, as one would
rather expect a scaling with the number of participants for soft
particle production. A scaling of the yield with the number of
participants would however lead to a value of $R_{AA} \approx 0.17$
for this centrality selection. This limit is not reached for the
lowest momenta covered ($p_{T} \approx 1 \, \mathrm{GeV}/c$), which
is most likely due to the broadening of the spectra in Au+Au compared
to p+p, which will increase the yield at finite $p_{T}$ more strongly
than the total multiplicity. At SPS energies it was observed \cite{ccc5}
that $R_{AA}$ was increasing with $p_{T}$ reaching a value of 1
already at $p_{T} \approx 2 \, \mathrm{GeV}/c$ -- this increase is
most likely due to a combined influence of increasing importance of
hard scattering at higher $p_{T}$ with possible initial state
multiple scattering and other momentum broadening effects like
collective transverse flow. The new results at RHIC, however, show a 
continuous decrease of $R_{AA}$ for higher $p_{T}$. This result is 
not explained by theoretical calculations which are shown in
Fig.~\ref{fig2} for comparison \cite{ccc9,ccc10,ccc11}. Calculations 
without energy loss fail completely, the calculations with
energy loss are closer in magnitude to the experimental results, but 
can so far not explain the $p_{T}$ dependence.

It has been argued by the PHOBOS collaboration \cite{ccc11a} that the particle
yield even at high $p_{T}$ is compatible with participant scaling for
Au+Au collisions with $N_{part} > 65$. This scaling is in general
not seen in peripheral Au+Au data and for the transition from p+p to 
Au+Au, a closer look at data from all experiments also shows that
even the approximate scaling is only valid in limited $p_{T}$ regions.
So while the participant scaling of the total multiplicities may well
have a fundamental origin, this is not at all clear for the same 
(approximate) scaling at a fixed $p_{T}$. Still it is interesting to 
note that $R_{AA}$ from neutral pions as shown in Fig.~\ref{fig2}
is compatible with participant scaling for the highest $p_{T}$. 

It has been
discussed whether the emission of high $p_{T}$ hadrons is dominated by 
the surface of the reaction zone, and thus
$R_{AA}$ would be determined by the surface-to-volume ratio. This
in turn contains further uncertainties from the surface thickness assumed.
To obtain a realistic numerical value
of the scaling in this scenario one would have to take into account
the transverse edge of the reaction zone and
the varying density of the nuclear overlap. No precise value is
available at present.
Besides, it is noteworthy that even at relatively high $p_{T}$
hydrodynamic particle emission might contribute significantly to
hadron spectra \cite{ccc12,ccc13}. For a locally thermalized system, 
which is the necessary condition for hydrodynamics, a scaling with
the number of participants is very natural, so this might apply to
the hydrodynamic fraction of the emitted hadrons.
At the present moment, however, it is unclear, whether the observed
scaling with the number of participants is purely a coincidence, or
whether it can be related to a physical mechanism. In any case, even 
if this scaling holds, it does require a suppression mechanism of
hard scattered particles when compared to the expectation from p+p
collisions.

Similar analyses have been performed on charged hadron spectra. So
far no direct p+p reference measurement is available. PHENIX has scaled the 
measured neutral pion spectrum in p+p to account for non-pions. The
nuclear modification factor as displayed on the right hand side of
Fig.~\ref{fig2} is slightly larger than the one for
neutral pions, but both are compatible within errors. The other RHIC 
experiments have used a parameterization of UA1 data as their
reference for $p_{T} < 6 \, \mathrm{GeV}/c$. The curves included in
Fig.~\ref{fig2} show fits extracted from the presented data, 
the systematic errors,
which are not shown, are of similar size as for PHENIX. All results
show qualitatively similar features: $R_{AA}$ increases at low
$p_{T}$, has a maximum around
$p_{T} = 2 - 2.5 \, \mathrm{GeV}/c$ and decreases for higher $p_{T}$. 
There are considerable differences in the magnitude of the observed
suppression for the
different experiments, although probably not very significant in view
of the systematic errors. For BRAHMS the relatively 
small value of the number of collisions used in the calculation 
may lead to an overestimate of $R_{AA}$. Part of the discrepancies 
may be due to the different reference distributions used. 

This is in 
line with the observation, that the collision-normalized ratio of 
central to peripheral spectra is in much better agreement between
PHENIX \cite{ccc1} and STAR \cite{ccc2}. Such a ratio is one
possibility to investigate the centrality dependence of the observed 
suppression. The ratio central to peripheral has the advantage that
no p+p reference is needed, however, it suffers from the larger
uncertainty of the estimate of the number of collisions in peripheral
reactions. While also in this representation a suppression is
observed, the new features of the collisions at RHIC are not as
striking here. Similar ratios for SPS energies by the WA98
experiment \cite{ccc5} show no enhancement, but also a suppression, which is
qualitatively similar, though weaker than at RHIC. 

The centrality dependence of the neutral pion suppression has been
investigated by PHENIX \cite{ccc3}, 
there is a gradual transition
from no suppression in peripheral collision to a strong suppression
in very central without any indication of threshold effects.
The centrality dependence of the charged hadron spectra has been studied in 
detail by STAR \cite{ccc2} and PHENIX \cite{ccc1}. A significant
change in shape is observed when going from peripheral to central
collisions. This can be best illustrated by looking at the truncated 
$\langle p_{T} \rangle$ as obtained by PHENIX \cite{ccc1}: For 
$p_{T} < 2 \, \mathrm{GeV}/c$ the $\langle p_{T} \rangle$ 
increases with increasing
number of participants, just like the $p_{T}$-broadening known from
lower energies which may be attributed to multiple scattering or
collective flow, which would be expected to get more important for
larger systems. For $p_{T} > 2 \, \mathrm{GeV}/c$ the 
$\langle p_{T} \rangle$
decreases with increasing number of participants.

\section{FLAVOR COMPOSITION AT HIGH $\mathbf{P_T}$}

\begin{figure}[tb]
    \centering
    \includegraphics{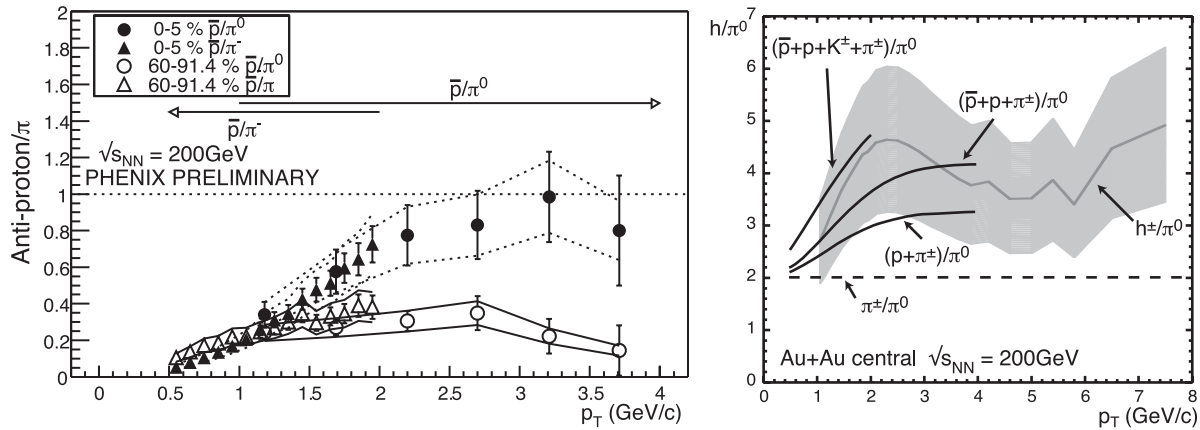}
    \caption{Left: Ratio of antiprotons to pions as a function of
    $p_{T}$ for peripheral and central Au+Au collisions at 
    $\sqrt{s_{NN}} = 200 \, \mathrm{GeV}$ 
    as measured by PHENIX \protect\cite{ddd1}. 
    Right: Ratios of identified charged hadrons to neutral pions in
    central Au+Au collisions at $\sqrt{s_{NN}} = 200 \, \mathrm{GeV}$. 
    The different curves show
    simple fits to the hadron ratios as presented by PHENIX \protect\cite{ddd1}.
    For comparison the ratio of non-identified charged hadrons to
    neutral pions in central collisions 
    is estimated from the presented results similar to 
    the same ratio for minimum bias collision as shown in
    \protect\cite{aaa7}.}
    \label{fig3}
\end{figure}

Differences in the suppression for charged hadrons and neutral pions 
may be partially due to a flavor dependence of the suppression
mechanism. Moreover, if the particle production at high $p_{T}$ is
actually due to incoherent hard scattering processes, the flavor
composition of hadronic spectra should approach values known from p+p 
collisions and should be calculable in pQCD. 
PHENIX has extended its previously 
reported baryon/pion ratios at 130 GeV to measurements at 200 GeV \cite{ddd1}. 
The same trend is observed: The proton- and antiproton-to-pion
ratios increase with increasing $p_{T}$ reaching values close to one 
at high $p_{T}$ in central collisions, as seen for antiprotons in
Fig.~\ref{fig3} on the left side. 
This is very different from the expectation of much 
smaller values seen in p+p and also in peripheral Au+Au reactions.
Other ratios, which may be calculated for hard processes, have been
looked at, like e.g. the antiproton/proton ratio, which is also
expected to decrease at high $p_{T}$ for hard scattering production.
PHENIX results are compatible with a constant, relatively large value
at high $p_{T}$ \cite{ddd1}, while STAR reports an indication of a
decrease at the highest $p_{T}$ \cite{aaa8}. 
But even taking this possible decrease into account, the
(anti-)proton/pion ratios indicate a still unexpected behavior in the
$p_{T}$ range, where charged hadrons can presently be identified. 

The strongest hint of the remaining puzzle comes again from the
neutral pion measurement in PHENIX. They have compared the neutral
pion yield to the charged hadron yield in minimum bias collisions
up to $p_{T} = 8 \, \mathrm{GeV}/c$ \cite{aaa7}, and report that even
at very high $p_{T}$ pions seem to make up only about $50 \% $ of the
total charged hadron yield. Other hadrons, possibly baryons, need to
contribute there in contrast to the expectations from p+p collisions. 
The right side of Fig.~\ref{fig3} tries to illustrate the current
situation. The charged hadron to neutral pion ratio has been
estimated from fits to the spectra presented by PHENIX for central
collisions. The identified charged hadron to pion ratios have then been 
used to successively build up this ratio for the total charged
hadrons. It can be seen that the total sum of pions, kaons, protons
and antiprotons is consistent with the charged hadrons up to
intermediate $p_{T}$. At high $p_{T}$ still a considerable
non-pionic contribution is apparently needed, and the measurement of 
identified hadron spectra at these higher $p_{T}$ will be of great
importance.

\section{ELLIPTIC FLOW}

\begin{figure}[tb]
    \centering
    \includegraphics{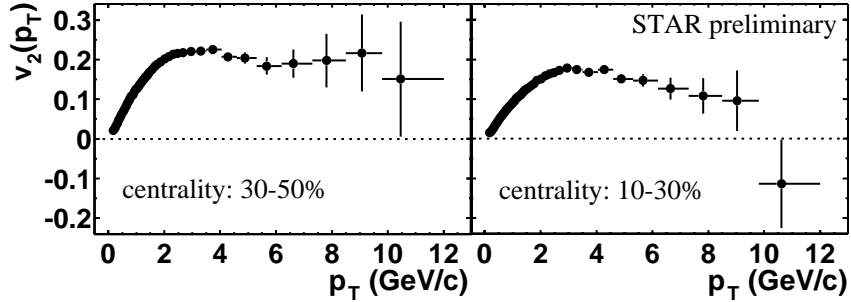}
    \caption{Elliptic flow parameter $v_{2}$ as a function of $p_{T}$ 
    for charged particles as measured by STAR
    \protect\cite{eee2} for two different semi-central samples of
    Au+Au collisions at 200~GeV. }
    \label{fig4a}
\end{figure}

A lot of new and improved results on elliptic flow have been
presented. A detailed summary is given in \cite{eee1}, so I will 
not discuss this in any detail. It should, however, be mentioned
that, 
while azimuthal asymmetries for low and intermediate $p_{T}$ appear
to be dominated by hydrodynamical effects, the persistence of a finite 
elliptic asymmetry $v_{2}$ out to very high $p_{T}$ indicates that modifications 
of hard scattering
production are likely relevant for these analyses.
This is nicely demonstrated in Fig.~\ref{fig4a}, which shows $v_{2}$ as
measured by STAR
\cite{eee2} out to beyond $p_{T} = 10 \, \mathrm{GeV}/c$. It will be
interesting to see at which momenta the asymmetry disappears, and how
much of the $v_{2}$ is actually due to partial jet quenching.

\section{JET STRUCTURES FROM AZIMUTHAL CORRELATIONS}

\begin{figure}[tb]
    \centering
    \includegraphics{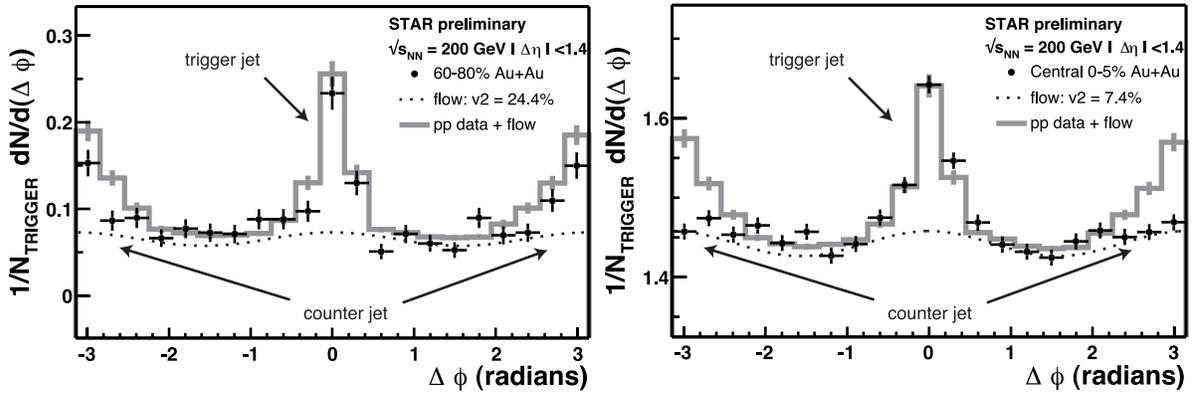}
    \caption{Azimuthal correlations of charged particles relative to 
    a high $p_{T}$ trigger particle as measured by STAR
    \protect\cite{fff1} for peripheral (left) and central (right)
    collisions. Estimates from p+p collisions are shown as grey
    histograms.}
    \label{fig4}
\end{figure}

While the suppression of high $p_{T}$ hadron production as seen in
inclusive spectra is certainly established and will yield important
constraints on the property of the hot and dense matter created in
these collisions, it is even more interesting to investigate jet
structures themselves. Analyses searching for jet patterns in
azimuthal correlations have been performed by STAR \cite{fff1,fff2} and PHENIX
\cite {fff3}. Both experiments see small angle correlations of
particles relative to high $p_{T}$ hadrons or photons, which in turn 
originate predominantly from pions. The magnitude and shape of this
correlation in Au+Au collisions is very similar to the one in p+p.
This is the first observation of jet structures in heavy ion
collisions. 

Also large angle correlations are seen by both experiments. STAR has 
performed a detailed analysis of these correlation phenomena
\cite{fff2}. As shown in Fig.~\ref{fig4} they attempt to describe the
correlation function in heavy ion collisions by a superposition of a 
flat background, a component of elliptic flow and the jet correlation
as measured in p+p. In peripheral collisions (left side of
Fig.~\ref{fig4}) both the trigger-jet ($\Delta \phi \approx 0$) and the
counter-jet ($| \Delta \phi | \approx \pi$) are very similar to the
expectations from p+p. In central collisions (right side of
Fig.~\ref{fig4}), the trigger-jet remains unaltered, but the
counter-jet has almost completely disappeared. The existence of the
trigger-jet ensures that a hard scattering has really taken place,
such that the suppression of the balancing counter-jet is a very
direct indication of jet quenching.

\begin{figure}[tb]
    \centering
    \includegraphics{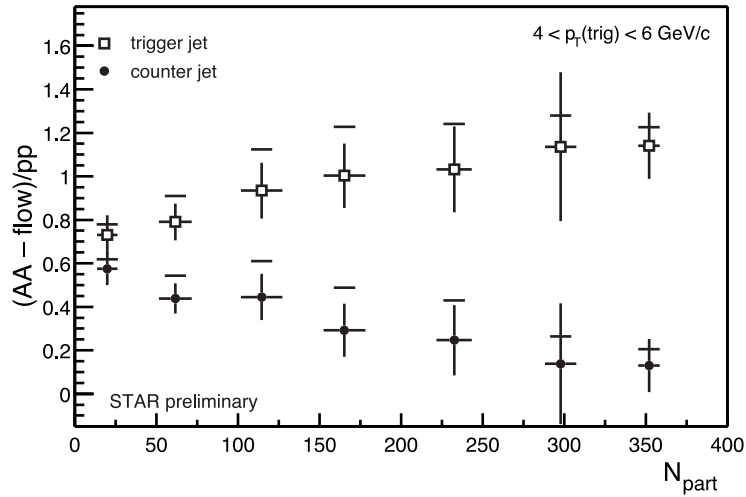}
    \caption{The relative strength of the trigger-jet (open symbols) and the 
    counter-jet (closed symbols) as a function of centrality in Au+Au
   collisions at 200 $A$GeV \protect\cite{fff1}.}
    \label{fig5}
\end{figure}

STAR has also presented the centrality dependence of the relative
strength of the two jet structures compared to p+p. As can be seen in
Fig.~\ref{fig5}, the strength of the trigger-jet shows even a small
increase towards more central collisions, which is surprising and
should be further investigated. The counter-jet strength exhibits a
gradual decrease with increasing number of participants,
qualitatively very similar to the centrality dependence of the high
$p_{T}$ neutral pion suppression. This measurements provides a very
important step towards understanding jet quenching, more detailed
systematic investigation of observed first trends in the data and
comparisons to theory are now required.

\section{SUMMARY}

A lot of exciting results on hard probes have been presented at this 
conference. The high $p_{T}$ hadron suppression in central Au+Au
collisions relative to p+p
collisions has been established, most notably through the neutral pion
measurements from PHENIX. Detailed theoretical calculations are now
eagerly awaited. Elliptic azimuthal asymmetries persist at very high 
$p_{T}$, which may be related to partial jet quenching.
Correlation structures due to jets have been
observed by STAR and PHENIX. The most important highlight of the
conference is the suppression of the counter-jet in triggered central
events as seen by STAR, which provides the most direct observation of
jet quenching to date. The most urgent open question concerns the
explanation of the large non-pionic contribution to hadron spectra at
very high $p_{T}$ as reported by PHENIX.

{\small \textit{
I would like to sincerely thank all speakers on high $p_{T}$ and
numerous other  
colleagues from the RHIC and SPS experiments for the help and support
in preparing my talk and these proceedings.} 
}


\begin{thebibliography}{9}
\bibitem{aaa1} M.~Gyulassy and M.~Pl\"umer, Phys. Lett. B \textbf{243}
(1990) 432; X.N.~Wang and M.~Gyulassy, Phys.Rev.Lett. \textbf{68} (1992)
1480.
\bibitem{aaa1a} R.~Baier, these proceedings.
\bibitem{aaa2} Proceedings of the Fifteenth International Conference on 
Ultra-Relativistic Nucleus-Nucleus Collisions (Quark Matter 2001), 
Nucl.Phys. A698 (2002).
\bibitem{aaa3} PHENIX collaboration, W.A.~Zajc, Nucl. Phys. A \textbf{698} 
(2002) 39c-53c; STAR collaboration, J.~Harris, 
Nucl. Phys. A \textbf{698} (2002) 64c-77c.
\bibitem{aaa4} PHENIX collaboration, K.~Adcox et al., 
Phys.Rev.Lett. \textbf{88} (2002) 022301.
\bibitem{aab4} STAR collaboration, C.~Adler et al., accepted in Phys.
Rev. Lett., preprint nucl-ex/0206011.
\bibitem{aaa5} BRAHMS collaboration, C.E.~J{\o}rgensen et al., these
proceedings.
\bibitem{aaa6} PHOBOS collaboration, C.~Roland et al., these
proceedings.
\bibitem{aaa7} PHENIX collaboration, S.~Mioduszewski et al., these
proceedings.
\bibitem{aaa8} STAR collaboration, G.J.~Kunde et al., these
proceedings.
\bibitem{bbb1} X.N.~Wang and M.~Gyulassy, Phys.Rev.Lett. \textbf{86} (2001) 3496; 
D.~Kharzeev and M.~Nardi, Phys.Lett. B \textbf{507} (2001) 121.
\bibitem{bbb2} PHOBOS collaboration, P.~Steinberg et al., these
proceedings.
\bibitem{ccc1} PHENIX collaboration, J.~Jia et al., these
proceedings.
\bibitem{ccc2} STAR collaboration, J.L.~Klay et al., these
proceedings.
\bibitem{ccc3} PHENIX collaboration, D.~d'Enterria et al., these
proceedings.
\bibitem{ccc4} D.~Antreasyan et al., Phys. Rev D \textbf{19} (1979)
764.
\bibitem{ccc5} WA98 collaboration, M.M~Aggarwal et al., Eur. Phys. J.
C \textbf{23} (2002) 225.
\bibitem{ccc6} UA1 collaboration, C.~Albajar  et al., Nucl. Phys. B \textbf{335}
(1990) 261.
\bibitem{ccc7} PHENIX collaboration, H.~Torii et al., these
proceedings.
\bibitem{ccc8} F.~Aversa et al., Nucl. Phys. B \textbf{327} (1989)
105.
\bibitem{ccc9} X.N.~Wang, Phys. Rev C \textbf{61} (2000) 064910.
\bibitem{ccc10} P.~Levai et al., Nucl. Phys. A \textbf{698} 
(2002) 631c.
\bibitem{ccc11} I.~Vitev et al., these proceedings.
\bibitem{ccc11a} PHOBOS collaboration, C.~Roland et al., these
proceedings.
\bibitem{ccc12} W.~Broniowski and W.~Florkowski, 
Phys.Rev.Lett. \textbf{87} (2001) 272302. 
\bibitem{ccc13} T.~Peitzmann, preprint nucl-th/0207012. 
\bibitem{ddd1} PHENIX collaboration, T.~Sakaguchi et al., these
proceedings.
\bibitem{eee1} S.~Voloshin, these proceedings.
\bibitem{eee2} STAR collaboration, K.~Filimonov et al., these proceedings.
\bibitem{fff1} STAR collaboration, C.~Adler et al., submitted to Phys.
Rev. Lett., preprint nucl-ex/0206006.
\bibitem{fff2} STAR collaboration, D.~Hardtke et al., these proceedings.
\bibitem{fff3} PHENIX collaboration, M.~Chiu et al., these proceedings.

\end{thebibliography}
\end{document}